# A quasi-bound band in the continuum in a photonic slab

Stanislav Tsoi,[1]* Nicholas Proscia,[1] Marc Christophersen,[1] Joseph Christodoulides,[1] Hsun-Jen Chuang,[1] Michael Povolotskyi,[2] Kathleen McCreary,[1] Paul Cunningham,[1] and Igor Vurgaftman[1]

[1]US Naval Research Laboratory, Washington, DC 20375, USA
[2]Amentum Solutions, Hanover, MD 21076, USA

**ABSTRACT**. Bound states in the continuum (BIC) are localized waves in electronic, photonic and acoustic systems, which remain decoupled from surrounding propagating waves and hence maintain their oscillation for extraordinary long time [Nat Rev Mater 1, 16048 (2016)]. In photonic crystals, symmetry-protected quasi-BICs (SP-qBIC) have been realized at high symmetry points of the Brillouin zone and utilized in photonic crystal and distributed feedback lasers. In the present work, we measure wavevector-resolved photoluminescence (PL) of monolayer $WSe_2$ weakly coupled to a photonic slab, consisting of a square array of aluminum nanodisks. The results show that the slab supports a continuous band of symmetry-protected quasi-bound states along the Γ-X direction, extending from the previously reported SP-qBIC at the Γ point. The spectral width of this quasi-bound band in the continuum remains narrow through at least a half of the Brillouin zone, indicating its long lifetime.

## I. INTRODUCTION.

The periodic variation of the dielectric function in photonic crystals shapes light properties in several fundamental ways [1]. The first is the formation of the photonic band structure – allowed modes are described by the dependence of the energy on the reduced wavevector within the first Brillouin zone [1]. Further, the periodic dielectric contrast acts as a diffraction grating, enabling Bragg scattering between degenerate photonic modes connected by the reciprocal vectors [2, 3]. This scattering means that such coupled modes do not exist separately, but always in pairs, triplets, etc. Consequently, the coexisting modes interfere and modify the photonic band structure predicted for non-interfering, independent modes. The interference is evident in the experimental observation of stop bands at high symmetry points in the Brillouin zone [4, 5], which do not occur in the independent mode picture.

The periodicity also dictates the symmetry of the modes resulting from the interference, with a crucial consequence that some modes become completely decoupled by the symmetry from the radiation continuum. These are the symmetry-protected bound states in the continuum (SP-BICs), possessing an infinite lifetime due to the radiative decoupling [6]. Whereas the SP-BIC with the infinite lifetime is a mathematical idealization, discrete SP-quasi-BICs (SP-qBICs) with a long but finite lifetime have been realized at isolated high symmetry points of the Brillouin zone and utilized in lasers [7, 8]. Recent works have claimed that strong coupling of SP-qBIC to light emitters can generate Bose-Einstein condensates of exciton-polaritons in photonic crystals [9, 10]. Given the technological and scientific impact of SP-qBICs, there is a fundamental need to understand better the periodicity-induced interference and symmetry in photonic crystals, in particular, whether interference generates bound states beyond the high symmetry points. In the present work, we use photoluminescence (PL) to examine the Brillouin zone of a photonic-crystal slab and demonstrate the existence of an intrinsic quasi-bound band in the continuum generated by interference.

## II. RESULTS AND DISCUSSION

The photonic-crystal slab used in this work consists of a square array of aluminum nanodisks fabricated on a flat fused silica substrate. The nanodisk diameter is near 100 nm, the height 20 nm, and the array period 520 nm. The rod-type slab [1] is formed when a 600 nm thick layer of poly-methyl methacrylate (PMMA), index-matched to silica, is spin coated on top of the array. To detect photonic modes of the slab in emission, the monolayer semiconductor emitter $WSe_2$ is mechanically transferred on top of the nanodisks prior to spin coating the PMMA layer.

Figure 1(a) shows wavevector-resolved transmission of the bare slab, recorded under the s-polarization (the electric field perpendicular to the measured in-plane wavevector). The spectrum contains two linearly dispersing branches of transverse-electric (TE) photonic modes propagating in the plane of the slab [1]. When $WSe_2$ is added to the slab, the same dispersing branches are observed in PL (Fig. 1b), indicating weak coupling between $WSe_2$ and the slab. Figure 1(c) explains schematically a two-step emission process leading to the detection of the photonic branches in PL. The photonic modes of the slab involve localized oscillations at each nanoparticle and are often referred to as the surface lattice resonance (SLR) [11, 12]. While a specialized theory has been developed for SLRs [13, 14], the present work employs the more general photonic crystal description [1], providing a more intuitive explanation of interference.

*Contact author: Stanislav.d.tsoi.civ@us.navy.mil

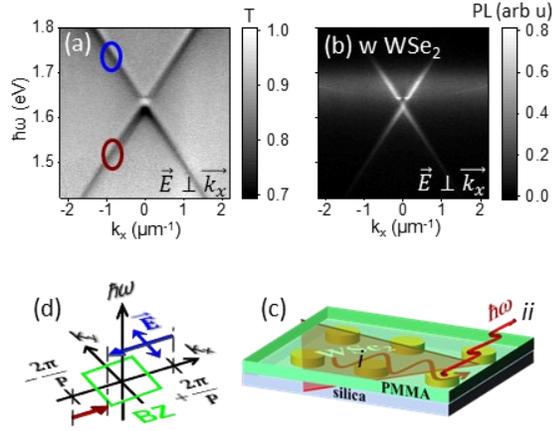

FIG 1. Dispersion of the photonic slab recorded in wavevector-resolved, s-polarized (a) transmission and (b) PL (with WSe$_2$). (c) Two-step PL process: i) the photo-excited WSe$_2$ excites a photonic mode and ii) the energy in the photonic mode scatters from the nanodisk array into far field. (d) Schematic representation of TE photonic modes, folded into the Brillouin zone (BZ) by the reciprocal vectors; $P$ is the slab period. The blue and brown arrows represent modes highlighted by blue and brown ovals, respectively, in the transmission spectrum.

Figure 1(d) presents schematically the photonic modes detected in transmission and PL, and shows that the observed linear dispersion is obtained due to folding the wavevector into the Brillouin zone by the reciprocal vectors $\pm 2\pi/P$ [1]. The modes composing the linear branches are the independent states, which do not experience interference.

The crossing point of the two linear branches at $\Gamma$ ($k_x$=0) represents an example of the photonic band structure modified by interference. The transmission and PL measurements show the formation of a stop band at $\Gamma$, with negligible extinction (Fig. 2a) and emission (Fig. 2b). The low-energy edge of the stop band appears broad and strong, whereas the high-energy edge narrow and weak, in agreement with previous measurements [10, 15, 16]. The general theory of photonic crystals attributes the origin of the stop band and contrasting edge states to interference of two degenerate counter-propagating modes (Fig. 2c), resulting in two standing waves (Fig. 2d, e) [1, 17, 18]. One of the standing waves has a symmetric distribution of the electric field with respect to the photonic crystal, with the neighboring anti-nodes occurring at the nanodisks and half-way between the nanodisks (Fig. 2d) [1]. The fields in the neighboring anti-nodes oscillate out of phase, and, when scattered by the polarizable material of the slab in the normal direction, produce destructively interfering waves. However, as the metallic nanodisks scatter light significantly stronger than the dielectric medium between them, the interference is only partially destructive and leads to strong emission into far field [18]. The emissive character of the symmetric standing wave represents a loss mechanism, limiting the lifetime of the mode and broadening its spectral extent. The lifetime is further reduced by ohmic losses in the metallic nanodisks. The experimental loss in the symmetric standing wave mode given by the full width at the half maximum (FWHM) is about 30 meV.

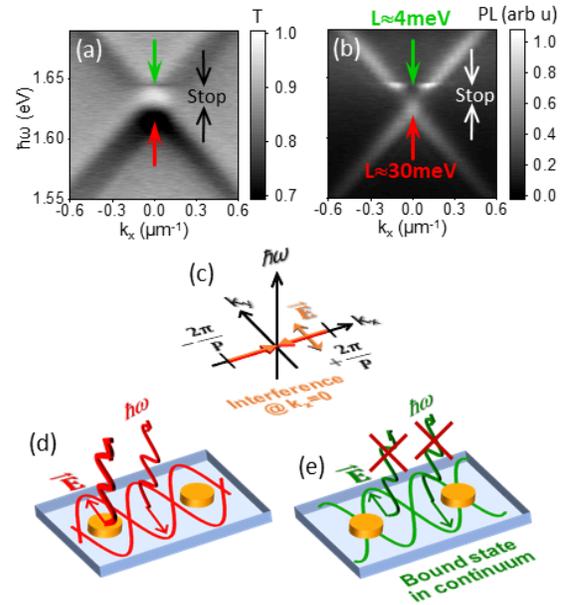

FIG 2. The s-polarized (a) transmission and (b) PL around the $\Gamma$ point ($k_x$=0). 'Stop' stands for the stop band at $\Gamma$. L is the loss of the edge states given by their FWHM. (c) Interference of two degenerate, counter-propagating modes at $\Gamma$. Schematic representations of resulting idealized standing waves with (d) symmetric and (e) anti-symmetric distributions of the electric field.

In contrast, the anti-nodes of the anti-symmetric standing wave experience equivalent dielectric backgrounds and, consequently, scattering of its energy into far field is suppressed fully due to the complete destructive interference (Fig. 2e) [18]. Further, compared to the symmetric standing wave, the anti-symmetric mode generates lower ohmic losses, as the nanodisks are aligned with the nodes. Due to the suppressed radiative and ohmic losses, the anti-symmetric standing wave can trap energy for a long time, earning the name of the symmetry-protected bound state in the continuum (SP-BIC) [6]. As the symmetric standing wave involves the strongly polarizable metallic nanodisks to a greater extent than the anti-symmetric wave, the frequency of the former is lower than the latter [1]. Thus, the strong and spectrally broad emission from the symmetric wave appears at the low-energy edge of the stop band, whereas the suppressed emission from SP-BIC occurs at the high-energy edge (red and green arrows in Fig. 2b, respectively). Numerical simulations confirm that the low- and high-energy edge states correspond to the symmetric and anti-symmetric standing waves, respectively (Supplemental Fig. S1).

*Contact author: Stanislav.d.tsoi.civ@us.navy.mil

The above discussion describes how interference generates the spectroscopic pattern characteristic of SP-BIC – the dark SP-BIC and bright state occupying the opposite edges of the stop band. Before we use this pattern to look for other bound states, however, let us address a prominent feature observed in PL, i.e. the two intense spots at small $k_x$'s on each side of the dark SP-BIC (Fig. 2b). They are occasionally observed in PL [19, 20] and typically detected in lasing, where they comprise most of the emitted light [9, 15, 21]. The ideal, infinitely extending SP-BIC results in perfect destructive interference of the scattered light (Fig. 2e), and therefore can generate neither spontaneous nor stimulated emission. However, in real photonic slabs all modes including standing waves have a finite size. Theoretical analysis indicates that the finite size relaxes the destructive interference of the scattered light at the end regions of the SP-BIC envelope, enabling some light emission into far field [22]. Spatially-resolved experiments have indeed confirmed that emission from the SP-BIC occurs primarily from the peripheral regions of the standing wave [16, 18]. The theoretical solutions also show that the periodicity of the standing wave varies across the SP-BIC envelope, matching the periodicity of the photonic crystal in the middle and deviating from it at the periphery [22]. Due to the periodicity mismatch, the emission from the peripheral regions emerges in the directions slightly off the normal, i.e. at small wavevectors (Fig. 2b). The emission via the side spots represents a loss mechanism, contributing together with the ohmic loss to the finite FWHM of the SP-BIC (Fig. 2b). Due to the finite losses, the SP-BIC has been lately referred to as a quasi-BIC (SP-qBIC) [23].

The above discussion emphasizes the existing detailed understanding of the origin of the SP-qBIC and of the characteristic PL pattern it generates – the stop band, the strong and broad low-energy edge, the suppressed emission at the high-energy edge and the intense side spots. We now exploit this characteristic PL pattern to look for quasi-bound states elsewhere in the Brillouin zone. Figure 3(a) reproduces the PL recorded along the $k_x$ axis, while Fig. 3(b) explains schematically how the experimental procedure selects the measured wavevectors from the Brillouin zone. An optical system (Supplemental Fig. S2) generates the Fourier transform of the emitting surface, i.e. the Brillouin zone, in the plane of the entrance slit of the spectrometer. The entrance slit then selects a single line from the Brillouin zone (Fig. 3b).

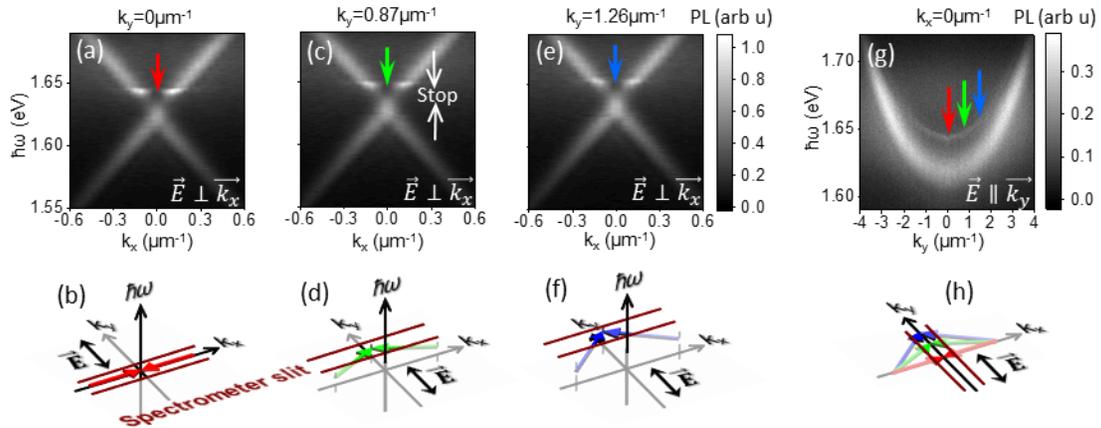

FIG 3. Wavevector-resolved s-polarized PL recorded along $k_x$ at (a) $k_y$=0, (c) $k_y$=0.87 µm$^{-1}$ and (e) $k_y$=1.26 µm$^{-1}$. (b, d, f) Schematic representations showing the entrance slit of the spectrometer selecting specific lines in the Brillouin zone for the $k_x$-resolved measurements, and interference of fully (b) and partially (d, f) counter-propagating modes. (g) $k_x$-resolved p-polarized PL showing that the low-energy bright and high-energy dark edge states merge into continuous lower bright and broad, and upper dark and narrow bands, respectively. (h) Schematic showing all interference modes captured simultaneously under the p-polarization. The vertical energy axis is omitted for simplicity.

*Contact author: Stanislav.d.tsoi.civ@us.navy.mil

Most previous measurements have restricted the position of the entrance slit to the high-symmetry axes, such as $k_x$ (Fig. 3b). In this work, we also start with the slit positioned at the $k_x$ axis (Fig. 3a, b), but subsequently translate the slit along the $k_y$ axis by 0.87 µm$^{-1}$, as depicted in Fig. 3(d). The PL spectrum recorded at the new position presents the familiar signatures of interference (Fig. 3c), suggesting that a bound-like state exists at ($k_x$=0, $k_y$=0.87 µm$^{-1}$). Figure 3(d) explains that the bound-like state results from interference of two partially counter-propagating modes, and therefore has a standing wave character along the x-axis, but propagating along the y-axis. The possibility of such partial interference in photonic slabs was proposed in an earlier work [24]. Numerical simulations confirm that the bound-like state has the anti-symmetric standing wave field distribution along the x-axis (Supplemental Fig. S3), consistent with the suppressed emission at $k_x$ = 0 and two bright side spots at small $k_x$'s (Fig. 3c). A further shift of the slit (Fig. 3f) reveals yet another bound-like state at ($k_x$ =0, $k_y$ =1.26 µm$^{-1}$) (Fig. 3e). Similar to the SP-qBIC at $\Gamma$, the bound-like states possess the narrow spectral width due to the symmetry-suppressed losses (Fig. 3c, e) and therefore constitute the quasi-bound states in the continuum. We note that all SP-qBICs in the present work are decoupled from the continuum by the symmetry and therefore distinctly different from accidental BICs [25, 26].

The obtained results suggest multiple SP-qBICs along the $k_y$ axis, which could be detected simultaneously if the entrance slit were aligned along the $k_y$ axis. The p-polarized PL recorded in such a geometry (Fig. 3h, polarization parallel to the measured wavevector) shows that the low-energy edge states merge into a continuous lower bright and broad band, whereas the dark high-energy edge states into an upper weak and narrow band, with a continuous stop band separating them (Fig. 3g). Being composed of the quasi-bound states, the upper band constitutes a *quasi-bound band in the continuum* (qBBC) along the $k_y$ axis. Another qBBC is expected along the $k_x$ axis due to the square-lattice symmetry of the slab.

A survey of existing literature suggests that the qBBC has been detected, but not properly interpreted, previously. Two parabolic bands were reported in the p-polarized PL of photonic slabs, composed of silver nanodisks [20, 27]. An investigation of a photonic crystal laser also reported two parabolic bands in unpolarized emission below the threshold [21]. The existence of a band with the quality factor approaching infinity is suggested in PL results from a 1D photonic crystal [9]. In addition, at least two studies have reported parabolic bands appearing to be the qBBC in transmission measurements [27, 28]. Furthermore, lasing from photonic-crystal slabs often features a parabolic dispersion extending significantly beyond the $\Gamma$ point, consistent with the narrow qBBC [27, 28]. We note, however, that these works neither commented on the physical origin of the band nor provided estimates of its spectral width. The original experimental demonstration of the SP-qBIC did report a very narrow band, but only in a very restricted range of wavevectors $k_x$<0.1 µm$^{-1}$ [4]. In addition, theoretical investigations predicted bands with a high Q-factor, but did not describe their physical origin [29-31].

The defining characteristic of bound states is their long lifetime or, equivalently, narrow spectral width. Figure 4(a) compares spectral widths of the qBBC and the low-energy bright band (Fig. 3g) to that of the non-interfering independent modes (Supplemental Fig. S5). The qBBC is significantly narrower than all other modes supported by the slab, indicating the increased lifetime due to the suppressed emission and lower ohmic loss. While a reliable width can be determined in the range $-2$ µm$^{-1}$ ≤ $k_x$ ≤ 2 µm$^{-1}$ (Supplemental Fig. S6), the qBBC remains narrow on visual inspection up to ±3 µm$^{-1}$ (Appendix I, Fig. 5a, b), i.e. through at least a half of the Brillouin zone.

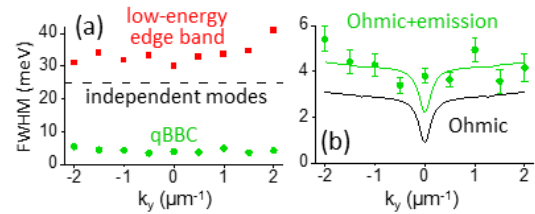

FIG 4. (a) FWHM of the qBBC, low-energy bright band, and independent modes. (b) The experimental FWHM of the qBBC (dots). The black curve is the numerical prediction for the ideal, infinitely extended qBBC (ohmic loss). The green curve is obtained by adding a constant radiative loss (1.3 meV).

The observed spectral width of the qBBC is between 3 and 6 meV, and numerical simulations suggest that most of the width is due to the ohmic loss in Al (Fig. 4b). The sub-meV SP-qBIC has been realized in lasing experiments using Ag, supporting a lower ohmic loss [5, 15, 16]. We attribute the difference between the experiment and simulations to the radiative loss from the qBBC, assumed constant in the lowest approximation (Supplemental Information).

We point that the resultant bright, narrow and dispersive qBBC is uniquely suited for strong coupling applications. Strong coupling between a photonic mode and light emitters generates mixed-character bosonic quasi-particles called exciton-polaritons [32]. The light effective mass of exciton-polaritons induced by their photonic component enables them to transition into the Bose-Einstein condensate (BEC) at significantly higher temperatures than the cold atom systems [32]. As a result, proposals to exploit polaritonic BEC for various applications have recently emerged [33-35]. We show that the qBBC offers the requisite characteristics for supporting the polaritonic BEC. First, the dispersion and narrow linewidth mean that the qBBC can generate dispersive polaritonic bands, similar to those realized using the dispersive Fabry-Perot band [32]. The dispersion of exciton-polaritons defines their effective mass and provides a continuous energy gradient for exciton-polaritons to thermalize and transition to BEC. Note that previous proposals for BEC in photonic crystals have considered only the isolated SP-qBIC at $\Gamma$ [9, 10], with unclear implications for the effective mass and thermalization mechanism.

*Contact author: Stanislav.d.tsoi.civ@us.navy.mil

Second, the optically bright character of the qBBC can be expected to produce emissive exciton-polaritons, facilitating non-destructive measurement of their distribution and condensate state [32]. Because the bright emission from the qBBC is achieved at the expense to the linewidth, it is critical that the band remains sufficiently narrow to support strong coupling. The criterion for strong coupling requires that the interaction between a photonic mode and light emitters exceeds their individual losses. Thus, the possibility of strong coupling is enhanced for the narrowest, linewidth-matched photonic modes and light emitters. Recently, narrow excitonic emission with the linewidth between 2 and 5 meV has been demonstrated in monolayer transition metal dichalcogenides (TMDs) encapsulated in multi-layer hexagonal boron nitride (hBN) [36, 37]. The present qBBC matches the linewidth of the hBN-encapsulated TMDs, therefore coupling the latter to a photonic-crystal slab could lead to optimal conditions for the polaritonic BEC.

### III. CONCLUSIONS

In summary, the present work demonstrates experimentally the existence of the quasi-bound band in the continuum in the photonic rod-like slab. The band is aligned with the $\Gamma$-$X$ axis of the slab and includes the previously reported symmetry-protected quasi-bound state in the continuum at the $\Gamma$ point. The band maintains the radiatively suppressed linewidth through at least a half of the Brillouin zone. A qualitative analysis of the band's origin attributes its enhanced lifetime to the symmetry-protected decoupling from the continuum. The narrow, bright and dispersive quasi-bound band provides optimal conditions for realization of the exciton-polaritonic BEC in photonic crystals.

### ACKNOWLEDGMENTS

This work was supported by the Office of Naval Research through the NRL Base Program and by the Office of Under Secretary of Defense for Research and Engineering through the Applied Research for the Advancement of S&T Priorities (ARAP) Program.

*Contact author: Stanislav.d.tsoi.civ@us.navy.mil

*Contact author: Stanislav.d.tsoi.civ@us.navy.mil

# APPENDIX I

The qBBC appears weak in Fig. 3(g), because the entrance slit is aligned with the $k_y$ axis and selects the dark edge of the quasi-bound states (Fig. 3a, c, e). While the high-energy edge appears dark (Fig. 3a, c, e), a close examination reveals weak, but non-zero emission (Supplemental Fig. S7), explaining the weak appearance of the upper band under the p-polarization. As each quasi-bound state also generates the bright side spots in PL (Fig. 3a, c, e), translating the entrance slit slightly off the $k_y$ axis to $k_x = -0.14$ µm$^{-1}$ reveals the qBBC in its highest emitted intensity (Fig. 5a); the second symmetric bright appearance of the qBBC at $k_x = 0.14$ µm$^{-1}$ is shown in Supplemental Fig. S8. The detection of the qBBC as two symmetric bright side bands surrounding the faint central band further supports its interference origin and quasi-bound character (Supplemental Fig. S8).

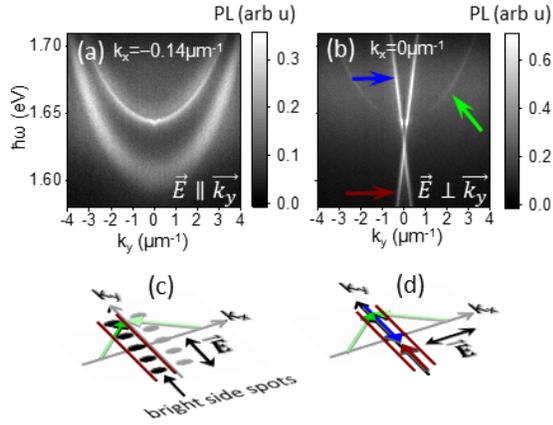

FIG 5. (a) Bright p-polarized PL emitted by the standing-wave component of the qBBC along $k_y$ at $k_x = -0.14$ µm$^{-1}$. (b) Weak s-polarized PL emitted by the propagating component of the qBBC along $k_y$ at $k_x = 0$. (c) Schematic representation of the spectrometer slit capturing the bright emission from the qBBC. (d) Partially counter-propagating modes impart the s-polarization to the qBBC (in addition to the p-polarization) enabling its detection in panel b.

While the standing-wave component of the qBBC is polarized along the *y*-direction (Fig. 5c) and emits p-polarized light (Fig. 5a), its propagating component is polarized along the *x*-direction (Fig. 5d) and enables the qBBC to simultaneously emit the s-polarized radiation (Fig. 5b). Note that the s-polarized spectrum also captures the linear dispersion of the orthogonal independent TE modes propagating along the *y*-axis (Fig. 5d). The s-polarized component of the qBBC is zero at $k_y=0$ where the propagating component disappears, but expected to become progressively stronger at greater $|k_y|$, as the angle between the two interfering modes decreases (Fig. 5d). We note that the s-polarized emission is intrinsic to the qBBC and expected to be present even in the ideal, infinitely extending mode. However, because most of the energy of the qBBC is contained in its p-polarized component (Supplemental Fig. S9), a contribution of this intrinsic loss to the linewidth remains minimal (Fig. 4B). Finally, the qBBC can also be detected in transmission (Supplemental Fig. S10).

The dispersion of the qBBC is similar to that of the resonant bands of the Fabry-Perot cavity. In the first approximation, the dispersion of the qBBC can be described by the relation

$$\hbar\omega = \frac{\hbar c}{n_{qBBC}}\sqrt{\frac{4\pi^2}{P^2} + k_y^2}, \qquad (1)$$

where, $c$ is the speed of light in vacuum and $n_{qBBC}$ an effective index of refraction. Compare this with the dispersion of the second band in the planar Fabry-Perot cavity:

$$\hbar\omega = \frac{\hbar c}{n}\sqrt{\frac{4\pi^2}{d^2} + k^2}, \qquad (2)$$

where $n$ is the refractive index inside the cavity, $d$ the distance between the mirrors, and $k$ the wavevector perpendicular to the axis of the cavity. At the zero wavevector, the cavity modes in the photonic slab and Fabry-Perot cavity are fully standing waves, and away from the $\Gamma$ point, the modes in both cavities are partially propagating waves. The main difference is that the Fabry-Perot bands exhibit the full rotational symmetry around the cavity axis, whereas the qBBC exists only along the symmetry axes of the photonic slab. In addition, the propagating component of the qBBC presents an intrinsic radiative loss mechanism, absent in the Fabry-Perot cavity.

*Contact author: Stanislav.d.tsoi.civ@us.navy.mil

SUPPLEMENTAL INFORMATION

**Sample Fabrication.** The photonic-crystal slab consists of a square lattice (250×250 μm$^2$) of aluminum nanodisks fabricated on a planar fused silica substrate (Fig. 1D). The circular nanodisk shapes with a 100 nm target diameter, arranged in a square lattice with the period of 520 nm, were defined in the electron beam photoresist ZEP using electron beam lithography. The patterns were developed in the ZEP developer for 60 seconds, rinsed in isopropyl alcohol (IPA), and dried under a nitrogen flow. The metal coatings with the thickness of 20 nm were deposited by electron beam evaporation, without a Ti adhesion layer. The subsequent lift-off was carried out in the Microposit 1165 remover at 80 °C, followed by the final rinse in IPA and acetone. The photonic slab was formed when a 600 nm thick layer of poly(methyl methacrylate) (PMMA) was spin coated on top of the nanodisk lattice. The matching indices of refraction of PMMA and silica created a homogeneous dielectric environment around the nanodisks, with the overall structure forming a rod-like photonic slab. To detect and measure the photonic states of the slab in emission, the monolayer semiconductor WSe$_2$ was mechanically transferred on top of the nanodisks prior to spin coating the PMMA layer (Fig. 1C).

**Measurements.** The wavevector-resolved transmission and PL were recorded using a custom-built optical setup represented schematically in Supplemental Fig. S2. For PL, continuous wave excitation at 631 nm and 300 μW from a semiconductor diode laser (Becker-Hickle BDU-SM series) was focused onto the WSe$_2$ flake from the silica side of the sample in a beam spot with the 30 μm diameter. The PL emitted by WSe$_2$ and the photonic slab from the PMMA side was collected by a 40X objective and the optical setup was used to generate the Fourier-transform of the slab (the Brillouin zone) at the plane of the entrance slit of the spectrometer (PI SpectraPro HRS 300), see Supplemental Fig. S2. The slit selected a single line from the Brillouin zone and the spectrometer grating dispersed the selected emission spectrally perpendicular to the slit, forming a 2D wavevector-resolved PL image on a Si CCD camera. For the transmission measurements, light from a broadband 3200K Halogen source was focused on a bare portion of the slab outside the WSe$_2$ flake into a circular spot with the diameter around 100 μm.

**Simulations.** Field distributions and ohmic losses of the modes of the photonic-crystal slab were calculated using the Wave Optics module of the finite-element package COMSOL (Version 6.3). The square array of 20-nm-thick Al nanodisks was embedded into a homogeneous medium with a refractive index of 1.48. Perfect magnetic boundary conditions were employed at the edges of a slab in order to isolate the in-plane-polarized modes interacting with the nanodisks. The dielectric permittivity of Al given in [1] was used without any adjustments. The diameter of the nanodisk $D$ and period of the array $P$ were adjusted to reproduce experimentally observed energies of the photonic modes. The adjustment is reflective of the uncertainty in the fabricated disk diameter and the precise refractive index of the surrounding medium (PMMA).

**Simulation results for the SP-qBIC at $\Gamma$.** Supplemental Figure S1 shows the electric field distributions for the low- and high-energy edge states of the stop band at the $\Gamma$ point, featuring the symmetric and anti-symmetric patterns, respectively. The calculations were performed for the ideal, infinitely extended states and included the ohmic losses only.

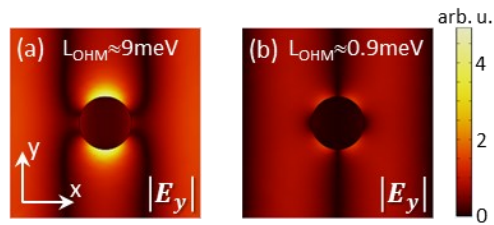

SUPP FIG S1. The electric field distribution of the (a) low- and (b) high-energy edge states of the stop band at the $\Gamma$ point calculated numerically. Ohmic losses are included.

**Optical system for wavevector-resolved measurements.** Supplemental Figure S2 shows schematically the optical system used to record the wavevector-resolved transmission and PL.

*Contact author: Stanislav.d.tsoi.civ@us.navy.mil

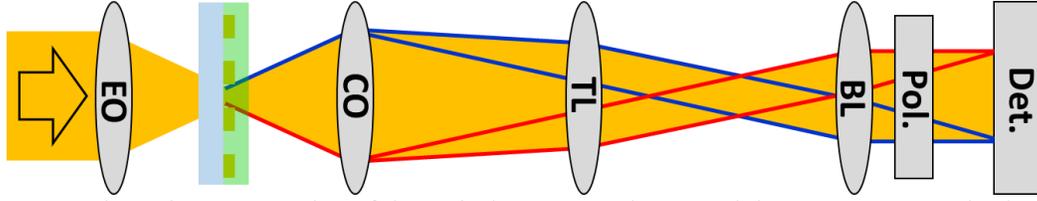

SUPP FIG S2. Schematic representation of the optical system used to record the wavevector-resolved transmission and PL. Optical components (from the left) include the excitation objective (EO), the collection objective (CO), the tube lens (TL), the Bertrand lens (BL), the Glan-Thompson polarizer (Pol.). The position of the detector corresponds to the Fourier-transform plane containing the Brillouin zone of the sample. For the wavevector- and photon energy-resolved measurements, the entrance slit of the spectrometer is positioned in the Fourier-transform plane. To record PL images in the full wavevector space for $\hbar\omega$ = const, a CCD camera is positioned at the Fourier-transform plane and a tunable narrow band-pass filter placed in front of the camera.

**Simulation results for qBBC.** Supplemental Figure S3(A) shows schematically that the SP-qBIC at $\Gamma$ results from full interference of 4 degenerate independent modes. Decomposing the electric field distribution of SP-qBIC into orthogonal components reveals a superposition of two standing anti-symmetric waves along the $x$- and $y$-axes, both generating zero radiative loss (Supplemental Fig. S3B, C). Supplemental Figure S3(D) depicts interference of partially counter-propagating modes outside $\Gamma$, leading to a quasi-bound state with the standing wave component along the $x$-axis and the propagating component along the $y$-axis (Supplemental Fig. S3E, F). The former exhibits the anti-symmetric distribution of the field, suppressing its radiative and Ohmic losses. Supplemental Figure S4 compares the photonic mode dispersion observed experimentally to the numerical simulations. A good agreement is reached when $D = 134$ nm and $P = 509$ nm are used in the simulations to correct for the experimental uncertainties.

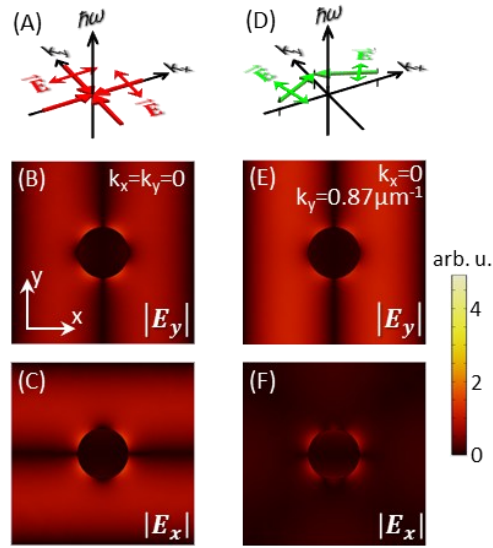

SUPP FIG S3. (A) Schematic representation of full interference among four degenerate modes at $\Gamma$. Double-headed arrows denote the electric field. The interference generates an SP-BIC displaying anti-symmetric electric field distributions along both (B) the $x$- and (C) the $y$-axes. (D) Interference outside $\Gamma$ involving two degenerate partially counter-propagating modes generates a quasi-bound state featuring (E) a standing wave with the anti-symmetric field distribution along the $x$-axis and (F) a propagating wave along the $y$-axis. The field distributions are calculated numerically for the infinite photonic slab.

*Contact author: Stanislav.d.tsoi.civ@us.navy.mil

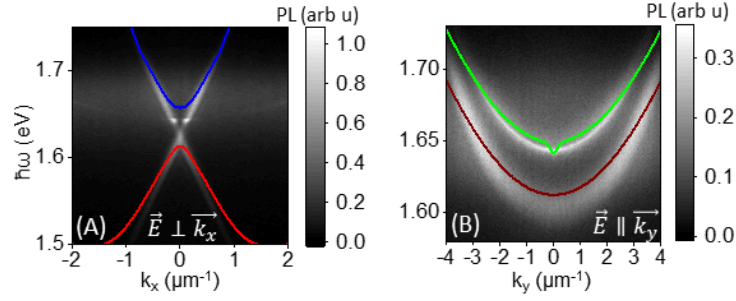

SUPP FIG S4. Comparison of the photonic dispersions observed experimentally and calculated numerically (color lines) in the (A) s- and (B) p-polarization.

**Spectral and wavevector widths of photonic modes.** Supplemental Figure S5 shows that the spectral and wavevector widths of the independent modes can be assessed from cross-sections of the s-polarized PL. To record the PL image in the full wavevector space at $\hbar\omega$=const (Supplemental Fig. S5D), a CCD camera is positioned at the Fourier image of the sample (Supplemental Fig. S2) and a tunable narrow band-pass (~3 meV) filter placed in front of the camera. The spectral full width at the half maximum (FWHM) of the independent modes is around 25 meV (Supplemental Fig. S5B). The longitudinal extent of the independent modes $\varDelta L$ is calculated using the uncertainty principle, $\Delta k \cdot \Delta L \sim 1$, where $\varDelta k$ is the FWHM in the wavevector extent of a monochromatic mode (Supplemental Fig. S5E). With the experimental $\Delta k \approx 0.12$ μm$^{-1}$, $\Delta L \sim 8$ μm. Supplemental Figure S5(F) explains that the circular contours observed in the PL image (Supplemental Fig. S5D) are formed when the TE modes with the constant energy $\hbar\omega$ propagating in different directions in the plane of the slab are folded into the Brillouin zone by the reciprocal vectors with the magnitude $2\pi/P$.

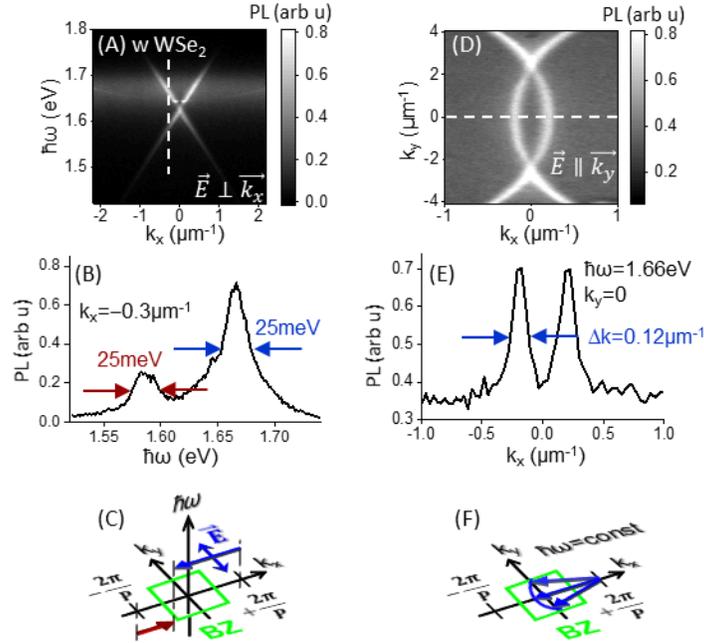

SUPP FIG S5. (A) The $k_x$-resolved s-polarized PL of the photonic slab with incorporated WSe$_2$ recorded at $k_y$=0. (B) The cross-section of the PL spectrum along the photon energy axis taken at $k_x$ = −0.3 μm$^{-1}$ (the dashed line in panel A). (C) Schematic representation of two independent photonic modes at a specific $k_x$ and $k_y$ = 0 in the Brillouin zone (BZ, the green square); $P$ is the slab period. The blue and brown arrows represent modes shown in panel B. (D) The PL image recorded in the full wavevector space for $\hbar\omega$ = 1.66 eV under the s-polarization. (E) The cross-section of the PL image along the $k_x$ axis (the dashed line in panel D). (F) Schematic representation showing independent modes with the energy $\hbar\omega$ (the blue arrows) forming one of the circular segments seen in panel D.

*Contact author: Stanislav.d.tsoi.civ@us.navy.mil

**Fits to the quasi-bound states.** Supplemental Figure S6 shows fits to PL cross-sections at specific $k_y$'s, performed for the bright appearance of the qBBC at $k_x = -0.14$ µm$^{-1}$ (Fig. 5).

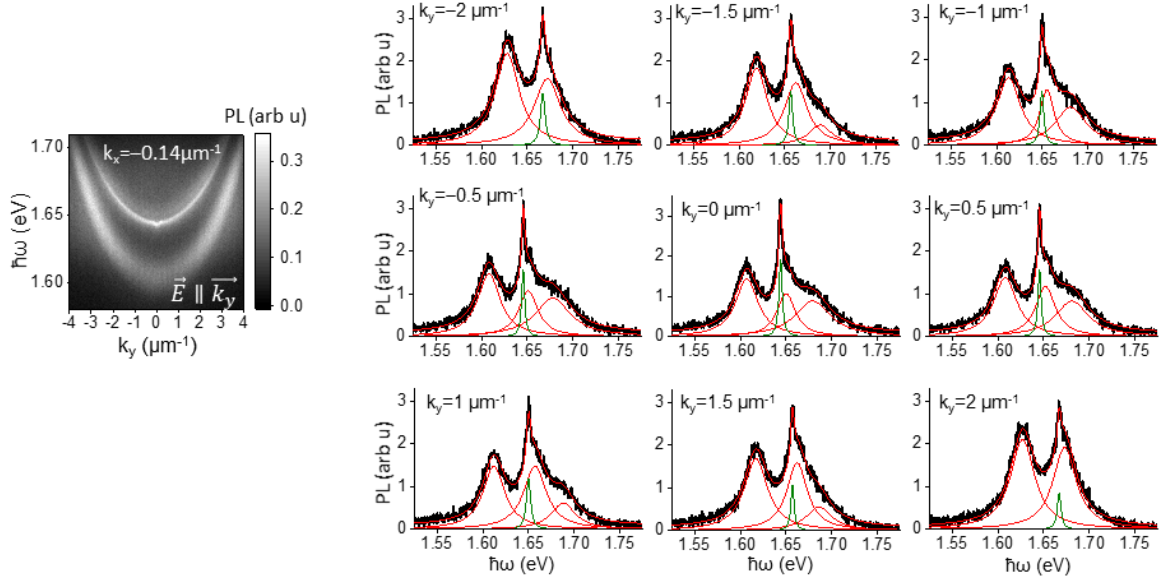

SUPP FIG S6. The $k_y$-resolved p-polarized PL recorded at $k_x = -0.14$ µm$^{-1}$ and fits to the cross-sections at specific $k_y$'s using a combination of 3 or 4 Lorentzian peaks. The narrow green Lorentzian corresponds to the qBBC.

**Additional PL and transmission results.** Cross-sections of the s-polarized PL taken at $k_x = 0$ reveal weak, but non-zero emission from the dark high-energy edge of the SP-qBICs (Supplemental Fig. S7).

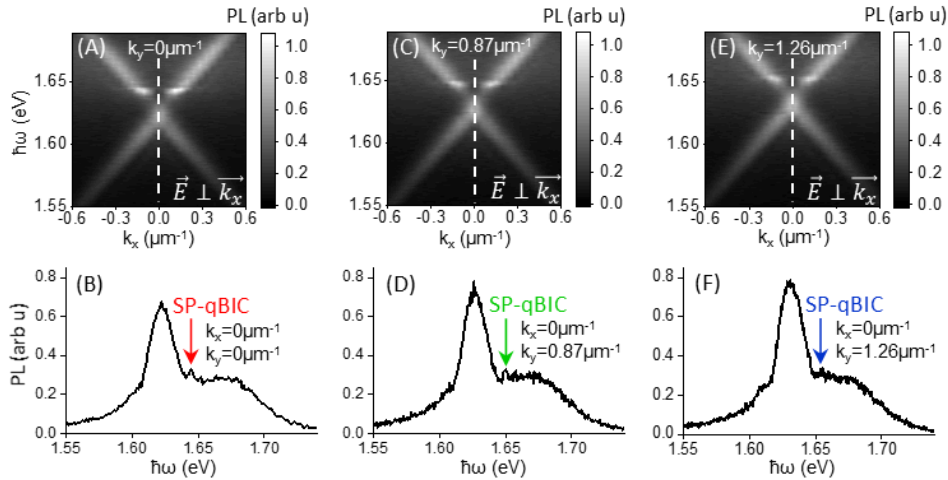

SUPP FIG S7. The $k_x$-resolved s-polarized PL of the photonic slab with incorporated WSe$_2$ recorded at (A) $k_y = 0$, (C) $k_y = 0.87$ µm$^{-1}$ and (E) $k_y = 1.26$ µm$^{-1}$. The dashed lines designate cross-sections along the photon energy axis taken at $k_x = 0$ and shown in (B, D, F), respectively. The red, green and blue arrows point to weak, but non-zero emission from the dark high-energy edge corresponding to the SP-qBICs.

Supplemental Figure S8(A-C) shows the $k_y$-resolved p-polarized PL recorded at $k_x = -0.14, 0, 0.14$ µm$^{-1}$, respectively, demonstrating that the qBBC appears weak at $k_x = 0$, and brightest at $k_x = -0.14$ and $0.14$ µm$^{-1}$. This strong appearance of the qBBC off the $k_y$-axis is clearly seen when spectra recorded at different $k_x$'s are combined into the full PL spectrum resolved along both $k_x$ and $k_y$ and the photon energy $\hbar\omega$ (Supplemental Fig. S8D). The broad emission seen at $k_y > 0.18$ µm$^{-1}$ is due to overlapping independent modes.

*Contact author: Stanislav.d.tsoi.civ@us.navy.mil

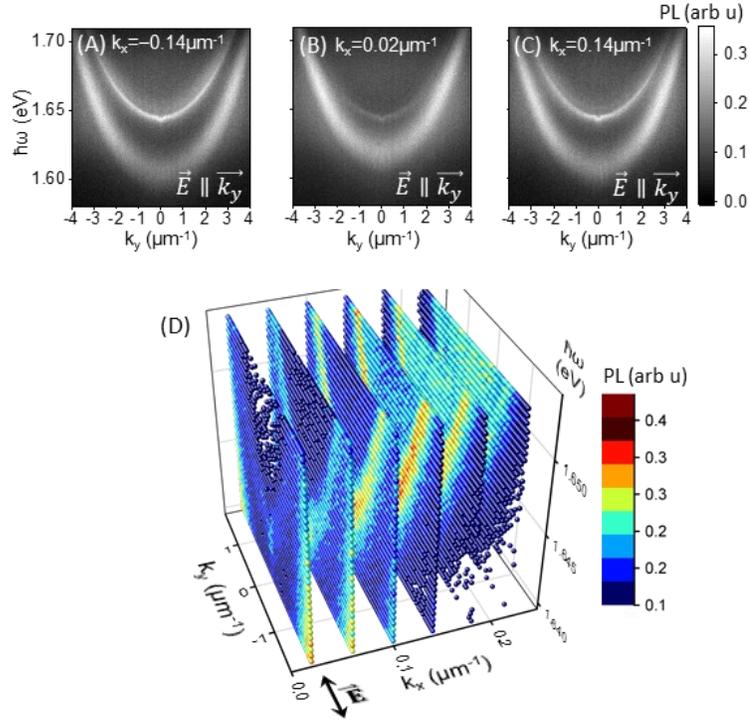

SUPP FIG S8. The $k_y$-resolved p-polarized PL recorded at (A) $k_x = -0.14$ μm$^{-1}$, (B) $k_x = 0.02$ μm$^{-1}$ and (C) $k_x = 0.14$ μm$^{-1}$. (D) The full PL spectrum resolved along $k_x$, $k_y$ and the photon energy $\hbar\omega$ showing the weak qBBC emission at $k_x = 0$ and strongest at $k_x = 0.14$ μm$^{-1}$. The symmetric part at negative $k_x$'s is omitted for clarity.

Supplemental Fig. S9 compares the p- and s-polarized PL spectra and shows that the standing-wave component of the qBBC emits significantly more light than the propagating component. The spectrometer response to the s-polarization is roughly twice stronger than to the p-polarization (Supplemental Fig. S9A, B), therefore the relative intensities of the two polarizations are adjusted to match the cross-sectional spectra below 1.55 eV and above 1.7 eV (Supplemental Fig. S9C). The amplitude of the qBBC in the p-polarization is roughly 3 times that in the s-polarization. Considering the two bright appearances of the qBBC in the p-polarization (Supplemental Fig. S8), the radiative loss in the standing-wave component exceeds that in the propagating component by at least a factor of 6. Further, in the investigated range of the wavevectors, the radiative loss by the standing-wave component of the qBBC can be assumed constant in the lowest approximation. Indeed, at $k_y = 2$ μm$^{-1}$, the interfering independent modes make a shallow azimuthal angle with the $k_x$-axis: $\varphi = tan^{-1}(k_y/[2\pi/P]) \approx 9.4°$ (Fig. 5d). Therefore, the standing-wave component ($\sim cos^2\varphi$) contains over 97% of the net energy of the qBBC.

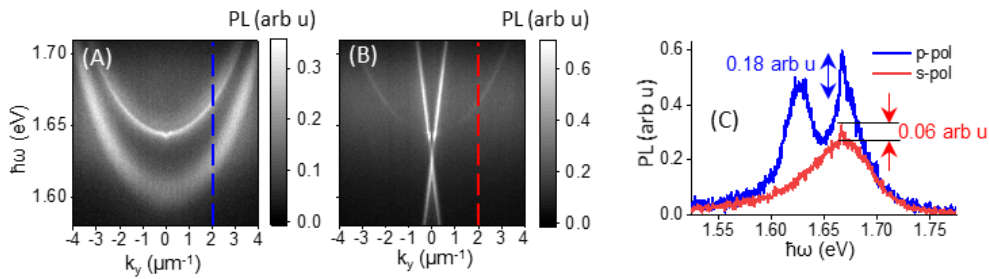

SUPP FIG S9. Wavevector-resolved PL of the photonic slab with incorporated WSe$_2$ recorded in the (A) p- and (B) s-polarizations. The blue and red dashed lines designate cross-sections taken along the photon energy axis at $k_y = 2$ μm$^{-1}$. (C) Comparison of the cross-sections showing relative intensities of the qBBC for the two polarizations of the emitted light.

*Contact author: Stanislav.d.tsoi.civ@us.navy.mil

Supplemental Figure S10 compares the wavevector-resolved p-polarized transmission and PL spectra recorded at $k_x = 0$. Both measurements reveal the low-energy bright and spectrally broad band, as well as the high-energy weak and narrow qBBC, with a continuous gap between them.

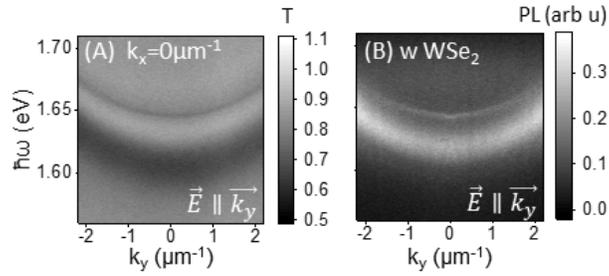

SUPP FIG S10. The $k_y$-resolved p-polarized (A) transmission and (B) PL recorded at $k_x = 0$.

*Contact author: Stanislav.d.tsoi.civ@us.navy.mil